\documentstyle[12pt,epsf]{article}

\advance \topmargin by -\headheight
\advance \topmargin by -\headsep

\evensidemargin \oddsidemargin
 \advance\hoffset by -3mm  
 \advance\voffset by  8mm  

\def\half{{{1\over 2}}}

\def\np{Nucl. Phys.}
\def\pl{Phys. Lett.}
\def\prl{Phys. Rev. Lett.}
\def\pr{Phys. Rev.}

\newcommand{\beq}{\begin{equation}}
\newcommand{\eeq}{\end{equation}}
\newcommand{\bear}{\begin{eqnarray}}
\newcommand{\eear}{\end{eqnarray}}

\newcommand{\LL}{{\cal L}}
\newcommand{\GG}{{\cal G}}
\newcommand{\HH}{{\cal H}}
\newcommand{\SF}{{\cal S}}
\newcommand{\tr}{{\rm Tr}}
\newcommand{\bo}{{\cal B}}

\newfont{\namefont}{cmr10}
\newfont{\addfont}{cmti7 scaled 1440}
\newfont{\headfontb}{cmbx10 scaled 1728}
\begin{document}
\begin{titlepage}
\begin{center} {\headfontb Physical States of Dyons}
\end{center}
\vskip 0.3truein
\begin{center}
{ M. Alvarez \footnote{e-mail: pyma@swansea.ac.uk}}
\end{center}
\vskip 0.2truein
\begin{center} {\addfont{Department of Physics,}}\\
{\addfont{University of Wales Swansea}}\\ {\addfont{Singleton Park, Swansea 
SA2 8PP, U.K.}}
\end{center}
\vskip 1truein

\begin{center}
\bf ABSTRACT
\end{center} 
It is shown that physical states of a non-abelian Yang-Mills-Higgs dyon are 
invariant under large gauge transformations that do not commute with its magnetic field. This result is established within an 
enlarged Hamiltonian formalism where surface terms are kept as dynamical variables. These additional variables are 
parameters of large gauge transformations, and are potential collective coordinates for the quantization of the monopole. 
Our result implies that there are no physical effects associated to some large gauge transformations and therefore their 
parameters should not be counted as collective coordinates. 
\vskip3.5truecm
\leftline{SWAT-97-156  \hfill June 1997}
\leftline{hep-th/9707117}
\vskip 1in
\begin{center}
{\bf \large Submitted to Physical Review Letters}
\end{center}
\smallskip
\end{titlepage}
\setcounter{footnote}{0}

\section{Introduction}
Monopoles in Yang-Mills-Higgs theories are known to posses a spectrum of excitations called dyons. In the 
simplest of these theories the gauge group is $\GG=SU(2)$ spontaneously broken to $\HH=U(1)$ by a non-zero vacuum
expectation value of the Higgs field. Gauss' Law implies that the system is invariant under small gauge 
transformations. Large gauge transformations with respect to the unbroken gauge group $U(1)$ induce an electric field
and therefore do have an observable effect; the monopole acquires electric charge and is then called a dyon. The gauge 
parameter of large $U(1)$ gauge transformations is thus interpreted as a collective coordinate of the monopole. 

The simple picture described in the preceding paragraphs is still valid when the gauge group $G$ is arbitrary but the Higgs 
field enforces maximal symmetry breaking. The unbroken gauge group is of the form $U(1)\times\ldots\times U(1)$ and for 
each $U(1)$ factor there will be an electric charge. In this paper we shall generalize the previous ideas to 
larger gauge groups and general symmetry breaking, paying special attention to transformation properties of physical states
under large gauge transformations. From the point of view of semiclassical
quantization, the most important question is to identify the internal degrees of freedom of a monopole when
the unbroken gauge group is non-abelian or, in more technical words, what is the moduli space of monopole solutions.
The parameters of the moduli space are identified with collective coordinates, and quantization of ``motion"
in moduli space yields a tower of dyon states. 

The example of maximal symmetry breaking suggests that the group parameters of the unbroken gauge group are collective
coordinates of the monopole, and that motion in the moduli space corresponds to large gauge transformations. However, it 
has been known for a long time that in a spontaneously broken Yang-Mills-Higgs theory not all generators of the unbroken 
gauge group $\HH$ correspond to collective coordinates in semiclassical quantization of monopoles If $\HH$ is non-abelian 
\cite{abou, nelson, cole, bala}. Problems arise when we try to define the action of the generators of $\HH$ that do not 
commute with the magnetic field of the monopole. We shall denote the set of these generators by $\HH'$. These problems 
can be exposed through semiclassical quantization \cite{abou} of monopole solutions, through topological considerations 
\cite{nelson}, or by studying the quantum mechanics of a test particle in the presence of a non-abelian monopole \cite{bala}.  
One of the aims of this paper is to analyze these problems from the point of view of physical states

It was shown in \cite{abou} that the momenta of inertia corresponding to $\HH'$ are vanishing, and that therefore the 
collective coordinates associated to those gauge generators somehow decouple from the theory. This fact suggests that
it should be possible to formulate the theory of non-abelian dyons in such a way that the 
decoupling of $\HH'$ is proven from first principles. In this paper we shall present a Hamiltonian formulation of non-abelian 
Yang-Mills-Higgs dyons that meets that 
demand. The crucial feature of a gauge theory is the appearance of Gauss' Law and the necessity of ``choosing a gauge" in 
order to obtain a non-singular symplectic structure acting on its phase space. Once the asymptotic values of the group 
parameters are given, the gauge condition fixes their value for all points. Therefore only the boundary values 
of the group parameters remain undetermined and can be considered dynamical variables \cite{wadia}. The extension of the 
Hamiltonian formalism to Yang-Mills-Higgs theories with dyons requires, therefore, the introduction of boundary terms that 
act as dynamical variables. These new variables enforce a new, extended Gauss' Law that will be shown to restrict physical 
states to singlets under $\HH'$. Therefore, there are no collective coordinates for the excitation of the $\HH'$ modes. The 
effect of the inclusion of a vacuum angle $\vartheta$ will also be considered.

The rest of this paper is organized as follows. Section 2 introduces the Hamiltonian formalism for Yang-Mills-Higgs theories, 
including the boundary terms that arise in the presence of dyons. Section 3 discusses the consequences of the extended
Gauss' Law on physical states. In Section 4 we present our conclusions.

\section{Boundary Terms in Yang-Mills-Higgs theories}
Let s consider a Yang-Mills-Higgs theory with simply-connected gauge group $\GG$ spontaneously broken to a subgroup 
$\HH$ by a Higgs field in the adjoint representation of $\GG$. We shall consider the theory to be defined on a large sphere 
$\SF$ with boundary $\bo$ in order to exhibit the importance of boundary terms in the action when the sphere contains 
magnetic monopoles. The lagrangean density of this theory is
\beq
\LL=\tr\left(-{1\over 4} F_{\mu\nu}\,F^{\mu\nu}-\half D_{\mu}\phi\, D^{\mu}\phi\right) -V(\phi),
\label{lagr}
\eeq
where the potential $V(\phi)$ must ensure spontaneous symmetry breaking. The classical action of this theory is, in 
first-order formalism,
\bear
S&=&\int_{t_1}^{t_2} dt\,\left[\int\,dx\,\tr\left(E_i \dot{A}_i + \pi\dot{\phi}\right)-H(E_i, A_i, \pi, \phi, A_0)\right],\nonumber\\
H&=&\int\,dx\,\tr\left[\half \,E_i\,E_i+\half \,\pi^2 +{1\over 4}\,F_{ij}\,F_{ij}+\half\,D_i\phi\,D_i\phi+V(\phi) 
-A_0\,\left(D_i\,E_i+e\,[\phi,\pi]\right)\right]\nonumber\\
&+&\int_{\bo} d\sigma_i\,\tr\,A_0\,E_i.
\label{act}
\eear
where the momenta are $E_i=F_{0i}$ and $\pi=D_0\phi$. Integrations in $x$ extend over the whole sphere
$\SF$. The surface element in $\bo$ is defined as $d\sigma_i=r^2\,\hat{r}_i\, d\omega$ with $\hat{r}_i$ a unit vector normal to
$\bo$ and $d\omega$ the solid angle element ($\omega$ is a shorthand for angular coordinates on $\bo$). The hamiltonian 
$H$ depends not only on the canonical coordinates and momenta, but also on $A_0$, the time 
component of the gauge potential. In standard expositions of gauge theories in the hamiltonian formalism it is assumed 
that the asymptotic value of $A_0$ is zero, so that when the radius of $\SF$ tends to infinity the last term in $H$ 
vanishes. Of course it is always possible to choose a gauge where $A_0$ vanishes at infinity, or even everywhere as in
the temporal gauge $A_0=0$. This possibility will be considered below, but for the sake of generality we shall keep the 
boundary value of $A_0$ arbitrary and possibly time-dependent \cite{wadia}. The spatial components of the 
gauge field $A_i^a$ are such that the monopole is a source of a magnetic field that, at large distances from the central
region of the monopole, takes the form
\beq
B_i={G(\omega)\over r^2}\,\hat{r}_i,
\label{B}
\eeq
with $G(\omega)=G^a(\omega)\,T_a$ an element of the unbroken gauge algebra $\HH$. The magnetic field must satisfy
the Bianchi identity, which implies that $G$ is covariantly constant, $D_i\,G=0$. At this point it is convenient to choose a 
Cartan basis for the generators of $\HH$, 
\bear
\HH&=&\left\{T_1, \ldots, T_l, T_{l+1},\ldots,T_r, E_{\alpha_1}, E_{-\alpha_1}, E_{\alpha_2}, E_{-\alpha_2},\ldots\right\}
\nonumber\\
\left[ T_I, E_{\alpha}\right] &=& \alpha_I\,E_{\alpha}, \\
\left[ E_{\alpha}, E_{-\alpha} \right] &=& \sum_{I=1}^l\alpha_I\, T_I , \nonumber\\
\left[ T_I, T_J\right]&=&0.
\label{alg}
\eear
The last $r-l$ generators $T_I$ correspond to possible abelian factors in $\HH$, while the first $l$ generate its maximal 
torus. The roots $\alpha$ are all non-zero, distinct, have non-vanishing components for $I=1,\ldots,l$, and span the 
$l$-dimensional space \cite{gno}. It is always possible to choose the generators of $\HH$ in such a way that
\bear
\tr(T_I\,T_J)&=&\delta_{IJ},\quad I,J=1,\ldots,l\nonumber\\
\tr(E_{\alpha}\,E_{-\alpha})&=&1,
\label{traces}
\eear
and the remaining traces of the products of two generators are zero. It will be convenient to work in the Wu-Yang formalism 
and use the freedom present in the definition of the maximal torus of $\HH$ to have 
\beq
G(\omega)=\sum_{I=1}^r g_I(\omega)\,T_I,
\label{Q}
\eeq
with $g_i$ the magnetic weights introduced in \cite{gno}. In this formalism the Higgs field is constant on $\bo$ and the 
unbroken gauge group $\HH$ lies inside $\GG$ in a position independent way. For future reference we will now give a 
criterion for the 
commutativity of a general element $X$ of $\HH$ with $G$. Let us write $X$ is the Cartan basis as
\beq
X=X_I\,T_I +X_{\alpha}\,E_{\alpha},
\label{x}
\eeq
where repeated indices $I$ are summed over. The commutator of $X$ with $G$ is easily found to be
\beq
[X, G]=\sum_{\alpha}(g\cdot \alpha)\,X_{\alpha}\,E_{\alpha}.
\label{xg}
\eeq
Therefore $X$ commutes with $G$ if and only if all the roots $\alpha$ included in the decomposition (\ref{x}) satisfy 
$g\cdot \alpha =0$. We shall denote those roots by $\alpha^{\perp}$: 
\beq
[X, G(P)]=0 \quad \Longleftrightarrow\quad X=X_I\,T_I + X_{\alpha^{\perp}}\,E_{\alpha^{\perp}}
\label{condition}
\eeq
Group elements will be represented by the parameters $\theta^a(x,t)$ that appear in the exponential map of the 
gauge group, $g=\exp(\theta^a\,T_a)\in\HH$. The asymptotic values 
of the group parameters $\theta^a(\omega,t)$ and their canonical momenta $P_a(\omega,t)$ will be included in the phase 
space of the theory. The motivation for this extension of the
phase space is that we shall impose Gauss' Law on physical states; this requirement, together with a choice of 
gauge, eliminates gauge transformations that vanish at infinity as symmetries of the theory but leaves the asymptotic
values of the group parameters undetermined. These asymptotic values are candidates for collective coordinates
of the dyon. As we want the dynamics of these degrees of freedom to appear as
equations of motion of the theory, we have to extend the phase space and the hamiltonian (\ref{act}) to accommodate the
new variables. We shall assume standard Poisson brackets for $\theta^a$ and $P_a$,
\bear
\{\theta^a(\omega,t), P_b(\omega',t)\}&=&\delta_b^{\,\,a}\delta(\omega-\omega')\nonumber\\
\{\theta^a(\omega,t), \theta^b(\omega',t)\}&=&\{P_a(\omega,t), P_b(\omega',t)\}=0.
\label{poi}
\eear
We shall consider physically static field configurations, that is those whose time dependence is given
by a gauge transformation:
\bear
\partial_0\,A_i&=&D_i\,\Lambda,\nonumber\\
\partial_0\,\phi&=&\left[ \phi, \Lambda\right],
\label{static}
\eear
where $\Lambda=\Lambda^a\,T_a$ is a gauge parameter that, at long distance from the monopole core, lies in the unbroken 
gauge algebra. It is always possible to write the asymptotic values of $\Lambda$ at any point $\omega$ in the boundary 
$\bo$ as 
\beq
\Lambda(\omega)=\Lambda_I(\omega)\,T_I+\lambda_{\alpha}(\omega)\,E_{\alpha}.
\label{lamb}
\eeq
We must also specify the boundary conditions that $A_i$ is assumed to obey. At large distance $A_i$ and $F_{\mu\nu}$ 
should vanish at least as fast as $r^{-1}$ and $r^{-2}$ respectively:
\beq
\lim_{r\to\infty}\,A_i = O(r^{-1}), \quad\quad\quad \lim_{r\to\infty}\,F_{\mu\nu}=O(r^{-2}). 
\label{cond}
\eeq
In addition we require that the radial component of $A_i$ decreases for large $r$ as $r^{-2}$,
\beq
\lim_{r\to\infty}\, \hat{r}^i\,A_i = O(r^{-2}).
\label{radial}
\eeq
These boundary conditions imply that $\hat{r}^i\,D_iA_0$ should decrease as $r^{-2}$. This 
fact does not imply that $A_0$ decreases as $r^{-1}$; it may behave at long distances like the 
Higgs field in the BPS limit, which has a non-vanishing limit ar $r$ goes to infinity and at the same time satisfies 
$D_i\phi=B_i=O(r^{-2})$. 

We still have to determine what is the time evolution of the asymptotic group parameters 
$\theta^a(\omega,t)$. We shall follow the assumption that their time evolution is a gauge transformation of parameter
$\Lambda=\Lambda^a\,T_a$, as in (\ref{static}). In order to express this idea we introduce of the Racah function 
$\Phi^a(\theta,\eta)$, which defines the product of two group elements:
\beq
\exp\left( i\,T_a\eta^a\right)\,\exp\left( i\,T_a\theta^a\right)=\exp\left( i\,T_a\Phi^a(\theta,\eta)\right).
\label{racah1}
\eeq
The derivative of this function with respect to its second variable acts as a vierbein in the sense that
gives the directions of small fluctuations about a certain point in the group manifold
\beq
E_b^{\,\,a}(\theta)={\partial\over\partial\eta^b}\,\Phi^a(\theta,\eta)\Big|_{\eta=0}.
\label{racah2}
\eeq
This vierbein relates elements of the Lie algebra to tangent vectors of the group manifold at a generic group element
$g$,
\beq
i\,T_a\,g=E_a^{\,\,b}\,\partial_b\,g,\quad\quad g=\exp(i\,\theta^a\,T_a)
\label{vier}
\eeq
where $\partial_a$ is the derivative with respect to $\theta^a$. The integrability condition of the Lie group requires 
$E_a^{\,\,b}$ to satisfy the following relationship
\beq
E_a^{\,\,c}\,\partial_c E_b^{\,\,d}-E_b^{\,\,c}\,\partial_c E_a^{\,\,d}=f_{ab}^{\,\,\,c}\,E_c^{\,\,d}.
\label{inte}
\eeq
The assumed time evolution of the group parameters of a general gauge group is therefore 
\bear
g(t+\delta t)&=&\exp\left\{i\Lambda\,\delta t\right\}\,g=\exp\left\{i\Lambda\,\delta t\right\}\,\exp\left\{i\,\theta^a\,T_a\right\}
=\exp\left\{i\,\Phi^a(\theta, \Lambda\,\delta t)\right\}\nonumber\\
&=&\exp\left\{i\,\left(\Phi^a(\theta,0)+E_b^{\,\,a}\,\Lambda^b\,\delta t\right)\right\}=
\exp\left\{i\,\left(\theta^a+E_b^{\,\,a}\,\Lambda^b\,\delta t\right)\right\}
\label{asu}
\eear
Writing the group element $g(t+\delta t)$ as $\exp(i\,\theta^a(t+\delta t)\,T_a)$ we conclude that the time derivative of
the parameter $\theta$ is
\beq
{d\over dt}\theta^a(\omega,t)= E_b^{\,\,a}(\theta)\,\Lambda^b(\omega,t),
\label{evol}
\eeq
Once we accept that the boundary values of the gauge 
parameters must evolve in time according to (\ref{evol}) and that the phase space of the theory must include the 
group parameters and their canonical momenta, it is necessary to extend the hamiltonian shown in (\ref{act}) with a new 
term that reproduces the equation of motion (\ref{evol}):
\bear
H'&=&\int\,dx\,\tr\left[\half \,E_i\,E_i+\half \,\pi^2 +{1\over 4}\,F_{ij}\,F_{ij}+\half\,D_i\phi\,D_i\phi+V(\phi) 
-A_0\,\left(D_i\,E_i+e\,[\phi,\pi]\right)\right]\nonumber\\
&+&\int_{\bo} d\sigma_i\,\tr\,(A_0\,E_i) +\int_{\bo} d\omega\, \tr\,(\Lambda\,J),
\label{ham}
\eear
where we have introduced the intrinsic momentum $J_a=E_a^{\,\,b}\/P_b$. The Poisson bracket of two $J$ reproduces the 
Lie algebra of the unbroken gauge group $\HH$,
\beq
\{J_a(\omega), J_b(\omega')\}=if_{ab}^{\,\,\,c}\,J_c(\omega)\,\delta(\omega-\omega'),
\label{lie}
\eeq
where $f_{ab}^{\,\,\,c}$ are the structure constants of $\HH$. We can think of the momenta $J_a$ as generators of $\HH$ and,
after quantization, as the operators that implement large gauge transformations on physical states. 

\section{Generalized Gauss' Law}
The last term of the extended hamiltonian $H'$ generates the desired time evolution for the group parameters through the
Poisson brackets (\ref{poi}). We are 
interested only in the equations of motion that, after quantization, are first order constraints on physical states of the
theory. These constraints should appear as stationary points of the action for variations of $A_0(x,t)$ and its boundary
values $A_0(\omega,t)$. The difficulty is that the variations of a field are not entirely independent of the variations of its
boundary values and therefore we cannot vary $A_0(x,t)$ and $A_0(\omega,t)$ independently. The way out of this 
problem is to restrict the phase space to a subspace where Gauss' Law is satisfied:
\beq
\left(D_i\,E_i +\left[ \phi, D_0\,\phi\right]\right)\,|\Psi\rangle =0.
\label{ga}
\eeq
where $|\Psi\rangle$ is a physical state. This eliminates $A_0(x,t)$ from the variational problem and leaves its boundary 
value $A_0(\omega,t)$ as the only lagrange multiplier. This is not the only effect of Gauss' Law; inserting the conditions 
(\ref{static}) into (\ref{ga}) leads to a link between $A_0$ and the gauge parameter that will be
most relevant to our discussion:
\beq
\left( D_i\,D_i + \left[\phi,\left[\phi,\quad\right]\right]\right)(\Lambda-A_0)=0.
\label{link}
\eeq
An obvious solution is $\Lambda=A_0$, which implies that the electric field is zero. It is therefore clear that 
$\Lambda-A_0$ must not vanish if we want to turn a monopole into a dyon. An important non-trivial solution is 
$\Lambda-A_0=C\,\phi$ with $C$ a constant; this solution, however, corresponds to picking the $U(1)$ direction in the 
unbroken gauge group defined by the Higgs field. It is easy to show that this would reproduce the usual charge quantization
condition on the Abelian electric charge of a dyon \cite{witten}. For that reason we will concentrate on the semisimple 
part of the unbroken gauge group, that is we shall ignore possible $U(1)$ factors in $\HH$. 

Far from the monopole core, (\ref{link}) is the Schr{\"o}dinger equation of a zero-energy adjoint particle in the presence of 
the magnetic field of the monopole. The long-distance behaviour of the solutions depends on whether the ``wave function'' 
commutes with $G$ or not \cite{abou}. We should therefore consider three cases:
\begin{enumerate}
\item Components of $\Lambda-A_0$ within the maximal torus clearly commute with $G$ as defined in (\ref{Q}). Equation 
(\ref{link}) reduces to Laplace's far from the monopole core, and thus the solution is of the form $(\Lambda-A_0)_I \sim 
Q_I\,r^{-1}$ with $Q\in\HH$ some $r$-independent operator. 
\item Components of $\Lambda-A_0$ corresponding to roots $\alpha^{\perp}$ defined by $\alpha^{\perp}_{\,I}\,g_{I}=0$ also 
commute with $G$ and thus behave as $(\Lambda-A_0)_{\alpha^{\perp}} \sim Q_{\alpha^{\perp}}\,r^{-1}$.
\item Components corresponding to the rest of the generators do not commute with $G$ and thus decrease faster,  
$(\Lambda-A_0)_{\alpha} \sim Q_{\alpha}\,r^{-n}$ with $n>1$.
\end{enumerate}
From now on we shall understand that roots $\alpha$ are not orthogonal to the vector whose components are the magnetic 
weights $g_I$ unless explicitly indicated by the index $\perp$. The general solution for $\Lambda-A_0$ is then
\beq
\Lambda-A_0 \sim {1\over r}\,Q_I\,T_I + {1\over r}\,Q_{\alpha^{\perp}}\,E_{\alpha^{\perp}}+{1\over r^n}\,Q_{\alpha}\,E_{\alpha},
\quad{\rm with}\quad n>1.
\label{so}
\eeq
It is clear that the asymptotic values $\Lambda(\omega,t)$ and $A_0(\omega,t)$ must coincide. This important fact
implies that variations of the boundary value of $A_0$ produce a new constraint that includes $J$:
\beq
\lim_{r\to\infty} r^2\hat{r}_i\,E_a^i+J_a=0.
\label{cons}
\eeq
The meaning of this new constraint is that large gauge transformations induce an electric field, thus giving electric charge to 
the monopole. Including the parameter $\Lambda$ of the gauge transformation, the quantum version of this constraint is
\beq
\left[\int_{\bo}d\sigma_i\,\tr\left(\Lambda\,E_i \right)+\int_{\bo}d\omega\,\tr(\Lambda\,J)\right]\,|\Psi\rangle=0.
\label{cons2}
\eeq
Together with (\ref{ga}), this constitutes a generalized Gauss Law \cite{wadia}. It is important to note that the radial part of 
the electric field should decrease exactly as $r^{-2}$ in order to contribute to the constraint (\ref{cons2}). Taking into 
account the general solution (\ref{so}), together with the fact that the radial component of $A_i$ already decreases as 
$r^{-2}$, we find that the radial part of the electric field behaves as
\beq
\hat{r}^i\,E_i =\hat{r}^i \,D_i\,(\Lambda-A_0) \sim {\partial\over\partial r}\,\left( {1\over r}\,Q_I\,T_I + {1\over 
r}\,Q_{\alpha^{\perp}}\,E_{\alpha^{\perp}}\right) +O(r^{-m}), \quad{\rm with}\quad m>2.
\label{oo}
\eeq
The implication of Eq. (\ref{oo}) is that the generators $E_{\alpha}$ disappear from the constraint (\ref{cons2}), due to the 
fact that the trace of $E_{\alpha}$ with $T_I$ or  $E_{\alpha^{\perp}}$ vanishes:
\beq
\int_{\bo}d\sigma_i\,\tr(\Lambda\,E_i)\sim -\int_{\bo}d\omega \left(\lambda_I\,Q_I + 
\lambda_{\alpha^{\perp}}\,Q_{-\alpha^{\perp}}\right).
\label{hot}
\eeq
We have used the traces given in (\ref{traces}). Before inserting the result (\ref{hot}) into the constraint (\ref{cons2}) we 
will introduce a by now obvious decomposition of the generators $J$:
\beq
J(\omega)= J_I(\omega)\,T_I+J_{\alpha^{\perp}}(\omega)\,E_{\alpha^{\perp}}+J_{\alpha}(\omega)\,E_{\alpha}.
\label{jota}
\eeq
Using (\ref{lamb}), (\ref{hot}) and (\ref{jota}) and comparing the terms in $\lambda_I$, $\lambda_{\alpha^{\perp}}$ and 
$\lambda_{\alpha}$ we finally find the action of the different components of $J$ on physical states:
\bear
J_I\,|\Psi\rangle &=& Q_I\,|\Psi\rangle,\nonumber\\
J_{\alpha^{\perp}}\,|\Psi\rangle &=&Q_{\alpha^{\perp}}\,|\Psi\rangle,\\
J_{\alpha}\,|\Psi\rangle &=&0.\nonumber
\label{result1}
\eear
We can summarize these results as follows: physical states transform under all the $T_I$ and under the generators 
$E_{\alpha^{\perp}}$. Large gauge transformations generated by the $E_{\alpha}$ leave physical states invariant and 
therefore do not correspond to collective coordinates. This phenomenon is due to the  anomalous behaviour of the radial 
part of the electric field in the isodirections $E_{\alpha}$.

\subsection{Inclusion of a vacuum angle}
The discussion of the preceding Section can be readily generalized to include a vacuum angle $\vartheta$. There is no need
to redo the calculations since the effect of the vacuum angle is essentially a shift in the electric field: 
\beq
E_i^a\rightarrow E_i^a-\vartheta\,B_i^a,
\label{angle}
\eeq
where $\vartheta$ absorbs some inessential constants. This can be plugged directly into Eq. (\ref{cons2}) with the following
results
\bear
J_I\,|\Psi\rangle &=& \left (Q_I+\vartheta\,g_I\right)\,|\Psi\rangle,\nonumber\\
J_{\alpha^{\perp}}\,|\Psi\rangle &=&Q_{\alpha^{\perp}}\,|\Psi\rangle,\\
J_{\alpha}\,|\Psi\rangle &=&0.\nonumber
\label{result2}
\eear
We see that physical states are invariant under the $J_{\alpha}$ even when a vacuum angle is introduced. The effect of 
the vacuum angle is limited to transformations under the commuting generators $J_I$. Equations (\ref{result1}) or 
(\ref{result2}) can be interpreted as a symmetry breaking induced by the non-abelian monopole whereby the unbroken 
gauge algebra $\HH$ is broken further down to its generators $T_I$ and $E_{\alpha^{\perp}}$. 

\section{Conclusions}
We have demonstrated that physical states of non-abelian dyons are invariant under large gauge transformations that do 
not commute with the magnetic field of the dyon. In the language of root systems those gauge transformations correspond 
to generators $E_{\alpha}$ of the unbroken gauge group with the root $\alpha$ not orthogonal to the magnetic weights. This 
result is still true if a vacuum angle $\vartheta$ is included. The only effect of the vacuum angle is an additional term in the 
transformation properties of physical states under the generators of the maximal torus of $\HH$. This additional term is 
proportional to the magnetic weights, which are weights of $\HH^v$, the dual of the unbroken gauge group \cite{gno}; 
physical states of dyons therefore carry a representation of $\HH^v$. This may be relevant to a better understanding of the 
Montonen-Olive conjecture \cite{mo}. 

Excitation of internal degrees of freedom of the monopole corresponding to the roots $\alpha$ not orthogonal to the 
magnetic weights induce no electric charge in the monopole and hence have no observable effects far from the monopole
core. From the point of view of collective coordinate quantization, this fact implies that there are no collective coordinates 
associated to those degrees of freedom. This result complements and clarify the results of references \cite{abou} to 
\cite{bala}.

\vskip 2cm
\begin{center}
\bf{Acknowledgements}
\end{center}
I would like to thank Professor T.J. Hollowood for many useful discussions.  This 
research has been supported by the British Engineering and Physical Sciences Research Council (EPSRC).

\vskip 2cm


\end{document}